\documentclass[a4paper,10pt, french]{article}
\usepackage[applemac]{inputenc}
\usepackage[T1]{fontenc}
\usepackage{lmodern,textcomp}
\usepackage{graphicx}
\usepackage{amsmath}
\usepackage[top=2.54cm, bottom=2.54cm, left=2.54cm, right=2.54cm]{geometry}
\usepackage[font=small,format=plain,labelfont=bf,up, indention=2.2cm, textfont = sl]{caption}
\usepackage{cite}
\usepackage{color}
\usepackage{parskip}
\usepackage{indentfirst}
\usepackage{float}
\usepackage{amssymb}
\usepackage[mathscr]{euscript}
\usepackage{cancel}
\usepackage{natbib}
\usepackage{scrextend}
\usepackage[caption=false]{subfig}
\usepackage{tikz}
\usetikzlibrary{arrows,shapes}
\newcommand{\kk}{\mathbf{k}}
\usepackage{soul}

\newcommand{\bjm}[1]{{{\color{green}#1}}}
\newcommand{\sean}[1]{{{\color{black}#1}}}

\begin{document}

\begin{center}

The role of parasitic modes in nonlinear closure via the resolvent feedback loop

Kevin Rosenberg, Sean Symon, and Beverley J. McKeon

\begin{addmargin}[3em]{3em}

We use the feedback loop of \citet{McKeon10}, where the nonlinear term in the Navier-Stokes equations is treated as an intrinsic forcing of the linear resolvent operator, to educe the structure of fluctuations in the range of scales (wavenumbers) where linear mechanisms are not active. In this region, the absence of dominant linear mechanisms is reflected in the lack of low-rank characteristics of the resolvent and in the disagreement between the structure of resolvent modes and actual flow features. To demonstrate the procedure, we choose low Reynolds number cylinder flow and the Couette equilibrium solution EQ1, which are representative of very low-rank flows dominated by one linear mechanism. The former is evolving in time, allowing us to compare resolvent modes with Dynamic Mode Decomposition (DMD) modes at the first and second harmonics of the shedding frequency. There is a match between the modes at the first harmonic but not at the second harmonic where there is no separation of the resolvent operator's singular values. We compute the self-interaction of the resolvent mode at the shedding frequency and illustrate its similarity to the nonlinear forcing of the second harmonic. When it is run through the resolvent operator, the `forced' resolvent mode shows better agreement with the DMD mode. A similar phenomenon is observed for the fundamental streamwise wavenumber of the EQ1 solution and its second harmonic. The importance of parasitic modes, labeled as such since they are driven by the amplified frequencies, is their contribution to the nonlinear forcing of the main amplification mechanisms as shown for the shedding mode, which has subtle discrepancies with its DMD counterpart.

\end{addmargin}

\end{center}

\section{Introduction}
The resolvent framework of \cite{McKeon10} can be used for low-order modeling \citep[e.g.][]{Rosenberg18b} and reconstruction of flows from limited measurements \citep[e.g.][]{Symon18b,gomez2016reduced}. Unlike data-driven methods such as dynamic mode decomposition DMD, which require snapshots of the fluctuating flow fields, only knowledge of the mean velocity profile is needed to form the resolvent operator which maps nonlinear terms, or forcing, to velocity fluctuations. One of the primary reasons for its success is the tendency for the resolvent operator to be low-rank at energetic frequencies. Under such conditions the optimal mode, which can be obtained via a singular value decomposition (SVD) of the operator, is significantly more amplified than its suboptimal counterparts and hence the flow structure can be approximated without precise knowledge of the nonlinear forcing. Effectively, then, mode structure is obtained by approximating the (action of the) resolvent operator.

\citet{Sharma16} and \citet{Towne17} have noted a correspondence between resolvent modes, DMD, and spectral Proper Orthogonal Decomposition (SPOD) modes. In particular, \cite{Towne17} noted
that resolvent modes are equivalent to SPOD modes when the nonlinear forcing is uncorrelated in space and time. When the nonlinear forcing is not white noise, resolvent modes deviate from their data-driven counterparts and this is particularly evident at frequencies where the amplification of the optimal mode is not sufficiently higher than the suboptimal modes.

The objective here is to take advantage of the feedback loop in \cite{McKeon13} and determine under what conditions mode shapes can be obtained by the action of the full resolvent on nonlinear forcing computed from resolvent modes, i.e. when it is appropriate to approximate the forcing rather than the resolvent, and how well these modes compare from data-driven approximations. In this respect, this study will share similarities to weakly nonlinear analyses \citep{herbert1983perturbation,Sipp07} in the sense of considering a limited set of finite-amplitude perturbations and their interactions. However, while weakly nonlinear analyses typically consider the neutral stability of a base flow and expansions in the vicinity of a critical Reynolds number, here we will consider the neutral stability of a 
mean flow and make no restriction \sean{on} the Reynolds number range. The flows analyzed herein are cylinder flow and the EQ1 equilibrium solution of Couette flow. Both flows exhibit strong amplification at a single frequency/spatial wavenumber which is linked to the neutral stability of the temporal mean profile in the case of the cylinder \citep{Barkley06} and the streamwise-averaged mean in the case of EQ1 \citep{hall2010streamwise}. \cite{MLugo14} leveraged this property to develop a self-consistent model for the cylinder where the Reynolds stresses are approximated by the most amplified mode whose amplitude is adjusted until the flow is neutrally stable. In a similar vein, the choice of EQ1 is motivated by previous work analyzing the emergence of streamwise-constant streaks, the scale interactions which give rise to them, and their connections to the self-sustaining process of turbulent shear flows \citep{waleffe1997self,hall2010streamwise,farrell2012dynamics}.


\section{Approach}

We consider the incompressible Navier-Stokes equations (NSE) and decompose the velocity field into a mean component $\overline{\boldsymbol{u}}$ and fluctuating component $\boldsymbol{u}'$. In the case of the cylinder, we use a temporal mean with $\overline{\boldsymbol{u}} = \left[\overline{u}(x,y),\overline{v}(x,y),0\right]^T$ and for EQ1 we use a temporal and streawise-averaged mean with $\overline{\boldsymbol{u}} = \left[\overline{u}(y,z),\overline{v}(y,z),\overline{w}(y,z)\right]^T$. The resulting equations for the fluctuations are
\begin{equation}
 \partial_t \boldsymbol{u}' + \overline{\boldsymbol{u}}\cdot \nabla \boldsymbol{u}' + \boldsymbol{u}' \cdot \nabla \overline{\boldsymbol{u}} + \nabla p' - Re^{-1} \nabla^2 \boldsymbol{u}' = -\boldsymbol{u}' \cdot \nabla \boldsymbol{u}' + \overline{\boldsymbol{u}' \cdot \nabla \boldsymbol{u}'}, \quad
  \nabla \cdot \boldsymbol{u}' = 0.
\label{eq:fluctuations}
\end{equation}

Following \cite{McKeon10}, we define $ \boldsymbol{f}' = -\boldsymbol{u}' \cdot \nabla \boldsymbol{u}' + \overline{\boldsymbol{u}' \cdot \nabla \boldsymbol{u}'}$ and treat $\boldsymbol{f}'$ as an unknown forcing. Upon Fourier transforming in the appropriate directions, we can recast equation \ref{eq:fluctuations} as
\begin{equation}
\hat{\boldsymbol{u}} = \boldsymbol{C}(i \omega \boldsymbol{I} - \boldsymbol{L})^{-1}\boldsymbol{B}\hat{\boldsymbol{f}} = \mathcal{H}(\kk) \hat{\boldsymbol{f}}, \label{eq:resolvent operator}
\end{equation}
where $\mathcal{H}(\kk)$ is the resolvent operator for a particular wavenumber/frequency vector $\boldsymbol{k}$, $\boldsymbol{L}$ is the linear Navier-Stokes operator associated with the mean, and $\boldsymbol{B},\boldsymbol{C}$ are operators which define the inputs and outputs; a complete mathematical definition of these operators for these flows is found in \cite{Symon18b}, \cite{Rosenberg18b}. It should be noted for the cylinder flow \sean{that} $\boldsymbol{k}$ is simply defined by the temporal frequency $\omega$ and in the case of the EQ1  by the streamwise wavenumber $k_x$ (and temporal frequency $\omega = 0$, as it is an equilibrium solution). Subsequently, we will analyze the resolvent operators $\mathcal{H}(\omega) $ and $\mathcal{H}(k_x)$ for the two flows respectively (where we have dropped the $\omega$ dependence in the latter). However, for generality and to facilitate comparisons between the flows going forward, we will maintain the notation $\mathcal{H}(\kk)$.

The resolvent operator relies on the mean velocity profile $\overline{\boldsymbol{u}}$ as an input and acts as a transfer function from nonlinearity $\hat{\boldsymbol{f}}(\kk)$ to velocity fluctuations $\hat{\boldsymbol{u}}(\kk)$ in Fourier space. A singular value decomposition (SVD) of the resolvent operator
\begin{equation}
\mathcal{H}(\kk) = \sum_p \hat{\boldsymbol{\psi}}_p(\kk)\sigma_p(\kk) \hat{\boldsymbol{\phi}}^{*}_p(\omega), \label{eq:resolvent SVD}
\end{equation}
leads to optimal sets of orthonormal singular response and forcing modes, or $\hat{\boldsymbol{\psi}}_p(\kk)$ and $\hat{\boldsymbol{\phi}}^{*}_p(\kk)$, respectively. The modes are ranked by the forcing-to-response gain $\sigma_p(\kk)$ under the $L_2$ norm. When the first singular value is much larger than the second one, i.e., $\sigma_1(\kk) \gg \sigma_2(\kk)$, the velocity response may be approximated as
\begin{equation}
\hat{\boldsymbol{u}}(\kk) \approx \hat{\boldsymbol{\psi}}_1(\kk)\sigma_1(\kk)\chi_1(\kk) = \hat{\boldsymbol{\psi}}_1(\kk)\sigma_1(\kk) \left<\hat{\boldsymbol{\phi}}_1(\kk),\hat{\boldsymbol{f}}(\kk)\right>. \label{eq:rank1}
\end{equation}
In cases where the singular values are not sufficiently separated and/or the forcing is highly structured, Equation \ref{eq:rank1} will not accurately reproduce the velocity fluctuations at a given $\kk$.

One course of action is to consider suboptimal resolvent modes ($p > 1$) and project the resolvent forcing modes onto the nonlinear forcing $\hat{\boldsymbol{f}}(\kk)$ to determine the complex weights $\chi_p$ associated with each resolvent response mode.
Here we approximate the structure of the fluctuations when the resolvent operator is not low-rank by considering the feedback loop introduced by \citet{McKeon13}, where the nonlinearity is formed by triadic interactions of the velocity modes, i.e.,
\begin{equation}
\hat{\boldsymbol{f}}(\kk_3) = \sum_{\kk_1 + \kk_2 = \kk_3}-\hat{\boldsymbol{u}}(\kk_1) \cdot \nabla \hat{\boldsymbol{u}} (\kk_2) \label{eq:NLF triad}
\end{equation}
The associated nonlinear forcing, consequently, is given by
triadically consistent resolvent response modes obtained from the previous singular value decompositions
\begin{equation}
\hat{\boldsymbol{f}}(\kk_3) = \sum_{\kk_1 + \kk_2 = \kk_3}\sum_a \sum_b -\sigma_a(\kk_1) \sigma_b(\kk_2) \chi_a(\kk_1) \chi_b(\kk_2) \left[\hat{\boldsymbol{\psi}}_a(\kk_1) \cdot \nabla \hat{\boldsymbol{\psi}}_b(\kk_2)\right].\label{eq:approximate NLF}
\end{equation}
All terms on the \sean{right-hand side} of Equation \ref{eq:approximate NLF} can be determined by approximation of the resolvent with the exception of $\chi_a$ and $\chi_b$, which may be obtained only under special circumstances, which we investigate here. In such cases, the forcing from Equation \ref{eq:approximate NLF} can then be used to obtain better approximations of the velocity fluctuations for the non-low rank case.

We compare the mode shapes arising from a rank-one approximation of the resolvent operator with those from approximation of the forcing for two flows\sean{. Both have} a strong physical mechanism underlying a dominant coherent structure in a narrow band of scales \sean{even though} energetic activity \sean{is} not limited to that band.

A direct numerical simulation (DNS) of two-dimensional cylinder flow is performed in FreeFem++ \citep{Hecht12} using the same mesh, boundary conditions, and discretizations as those in \citet{Symon18} at $Re = UD/\nu=100$, based on cylinder diameter $D$, inlet velocity $U$, and kinematic viscosity $\nu$. The time-step is $\Delta t = 0.02$ and the mean is obtained after time-averaging over 25 vortex shedding cycles once the flow achieves a steady limit cycle. The linear operators needed for resolvent analysis are formed in FreeFem++ around the two-dimensional mean flow, and resolvent modes are computed using the procedure outlined in \citet{Sipp13}. The DNS snapshots
are analyzed using DMD \citep[e.g][]{Schmid10, Rowley09} to isolate the true structure of the flow at individual frequencies.

The Couette equilibrium solution EQ1 \citep{nagata1990three,gibson2009equilibrium} is considered under the same domain and discretization as outlined in \cite{gibson2009equilibrium} with Reynolds number of $Re = 1000$.
 The discrete spatial operators needed to form the resolvent utilize Chebyshev and Fouier differentation matrices in the wall-normal and spanwise directions respectively \citep{weideman2000matlab}.

\section{Approximation of Cylinder Flow \& Exact Coherent State}

\begin{figure}
	\centering
	\subfloat[]{\includegraphics[scale=0.25,trim=0cm 0cm 0.0cm 0cm, clip=true]{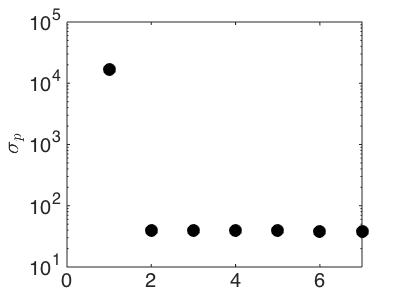}}
	\subfloat[]{\includegraphics[scale=0.25,trim=0.5cm 0cm 0.5cm 0cm, clip=true]{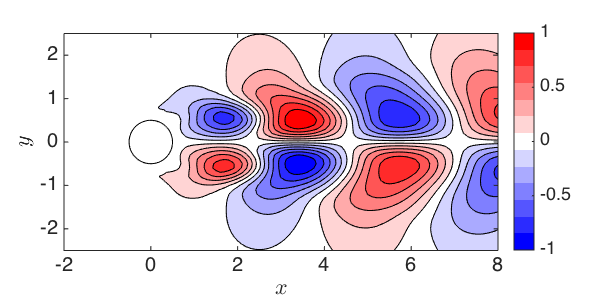}}
	\subfloat[]{\includegraphics[scale=0.25,trim=0.5cm 0cm 0.5cm 0cm, clip=true]{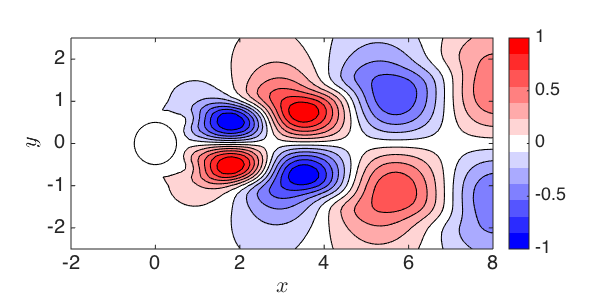}}
	
	\subfloat[]{\includegraphics[scale=0.25,trim=0cm 0cm 0.5cm 0cm, clip=true]{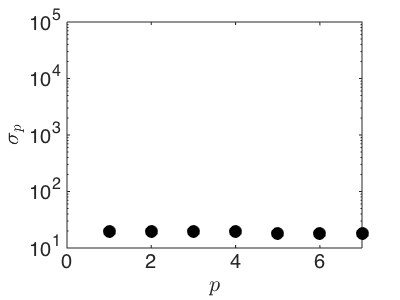}}
	\subfloat[]{\includegraphics[scale=0.25,trim=0.5cm 0cm 0.5cm 0cm, clip=true]{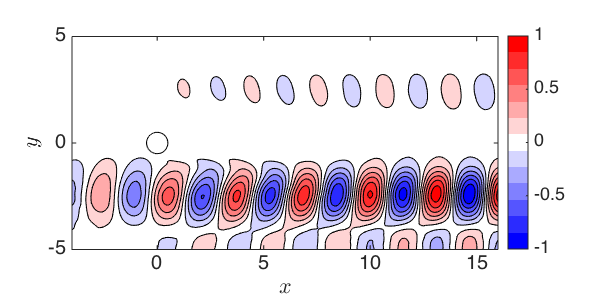}}
	\subfloat[]{\includegraphics[scale=0.25,trim=0.5cm 0cm 0.5cm 0cm, clip=true]{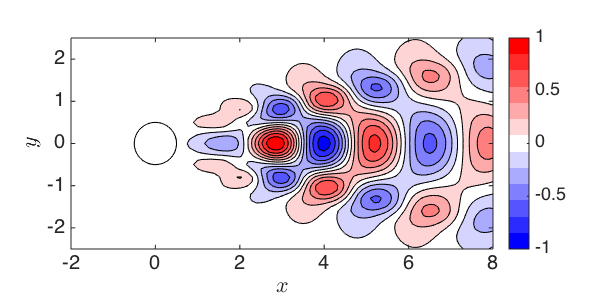}}
	
	\subfloat[]{\includegraphics[scale=0.22,trim=0.5cm 0cm 0.5cm 0cm, clip=true]{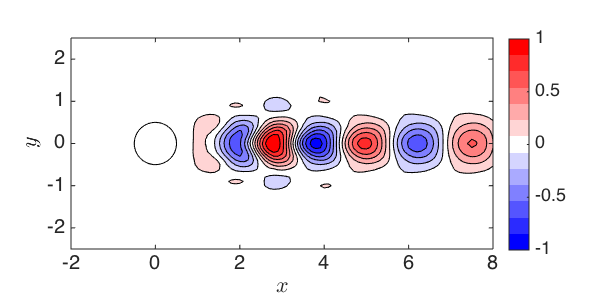}}
	\subfloat[]{\includegraphics[scale=0.22,trim=0.5cm 0cm 0.5cm 0cm, clip=true]{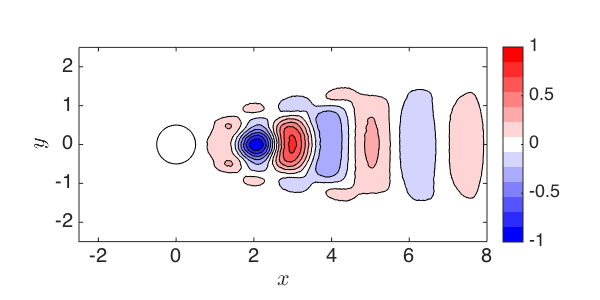}}
	\subfloat[]{\includegraphics[scale=0.22,trim=0.5cm 0cm 0.5cm 0cm, clip=true]{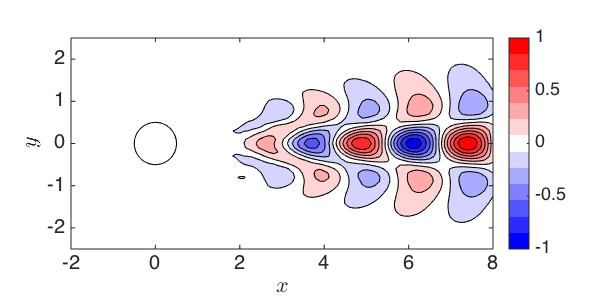}}
	
	\subfloat[]{\includegraphics[scale=0.32,trim=0cm 0cm 0.0cm 0cm, clip=true]{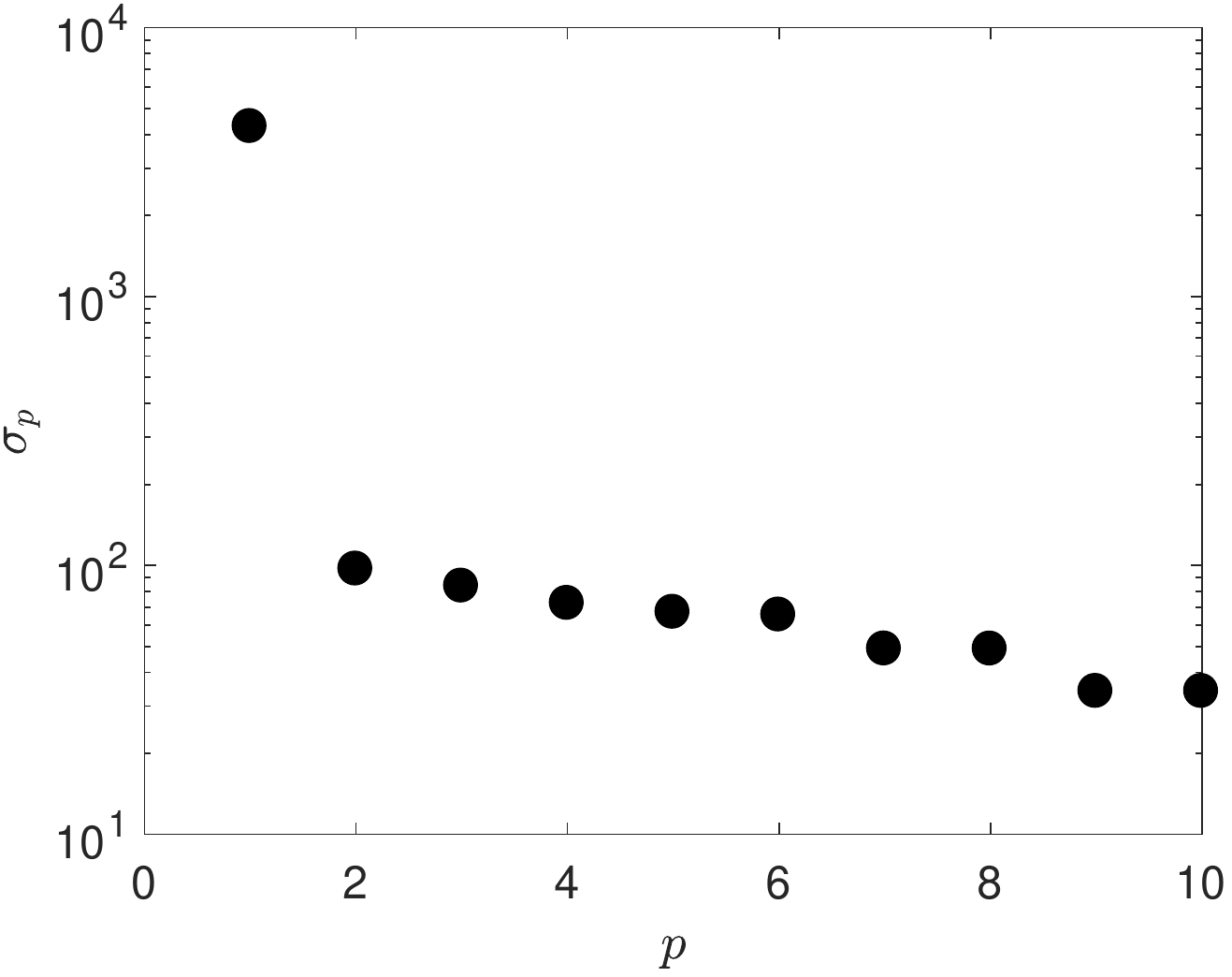}}
	\subfloat[]{\includegraphics[scale=0.32,trim=0.0cm 0cm 0.0cm 0cm, clip=true]{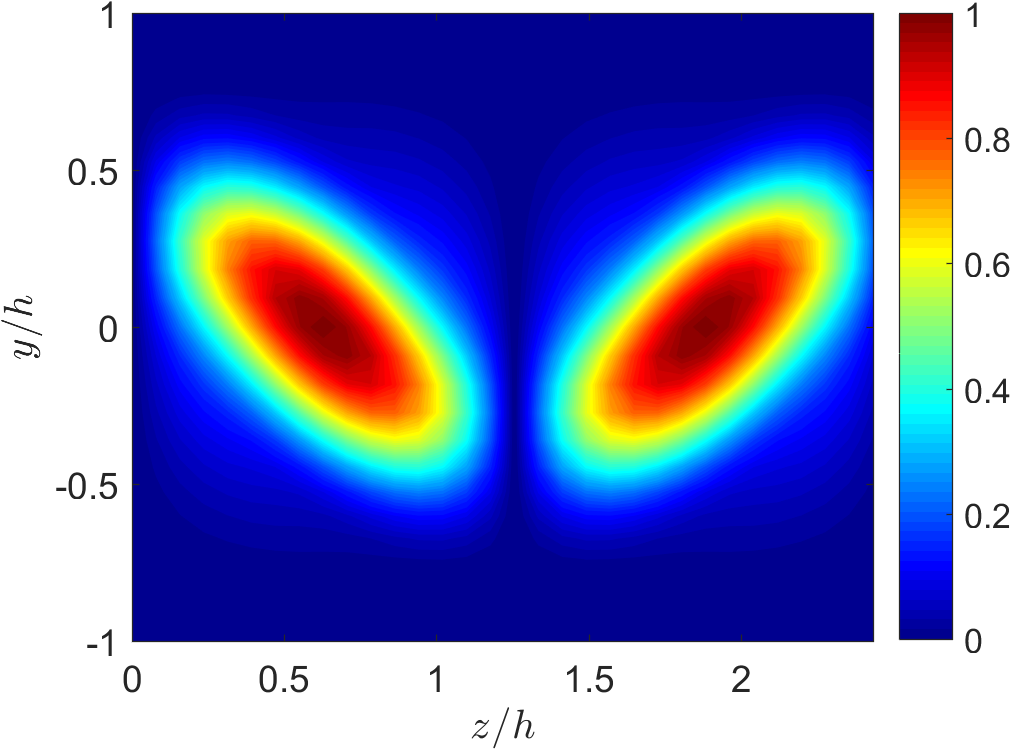}}
	\subfloat[]{\includegraphics[scale=0.32,trim=0.0cm 0cm 0.0cm 0cm, clip=true]{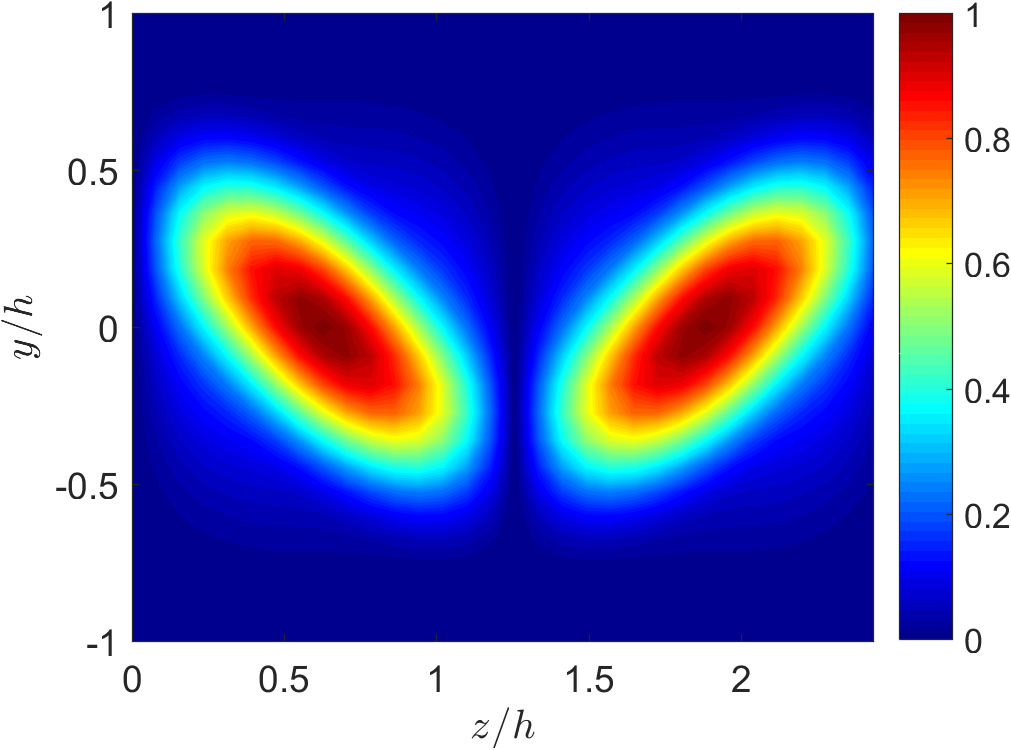}}
	
	\subfloat[]{\includegraphics[scale=0.32,trim=0cm 0cm 0.0cm 0cm, clip=true]{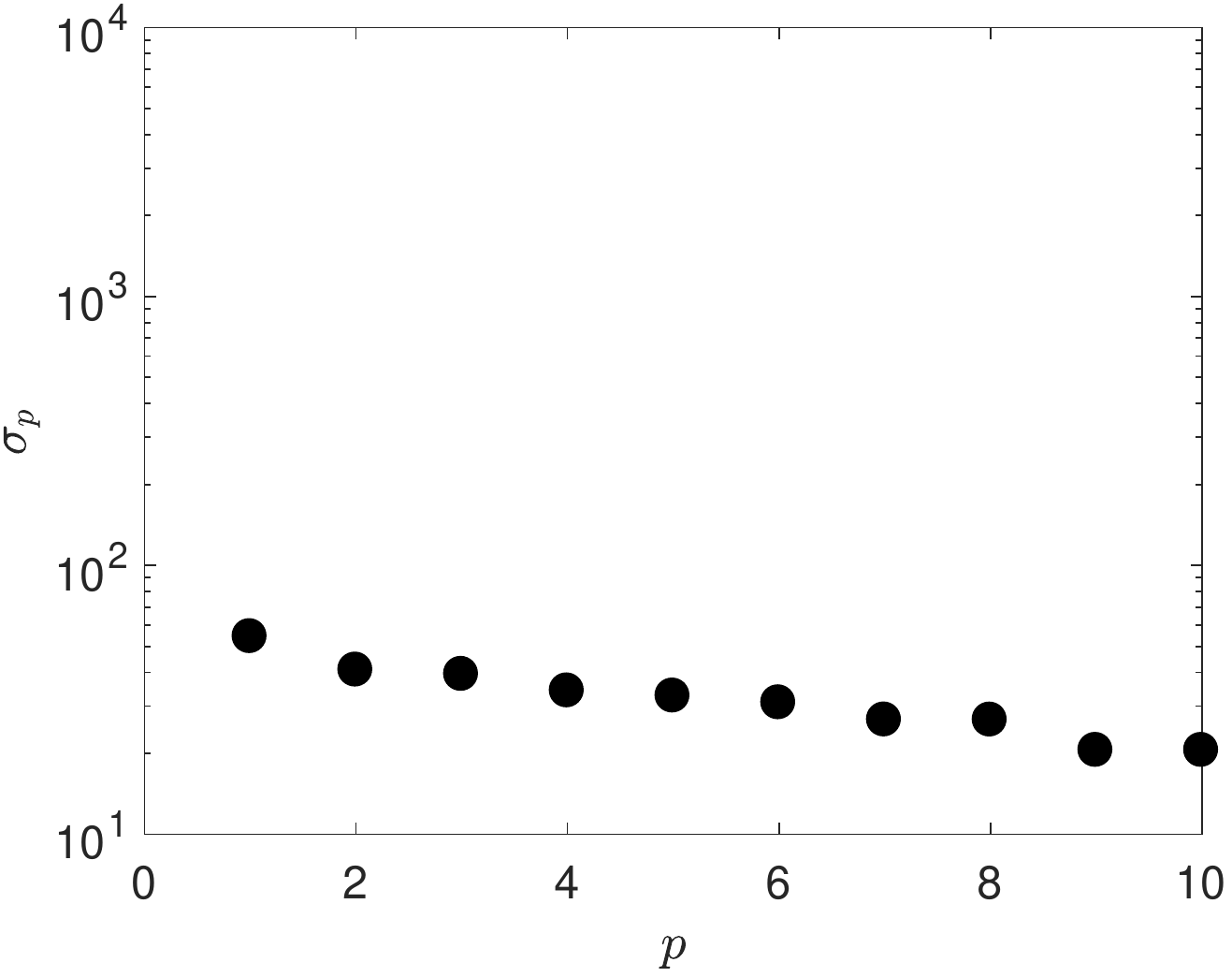}}
	\subfloat[]{\includegraphics[scale=0.32,trim=0.0cm 0cm 0.0cm 0cm, clip=true]{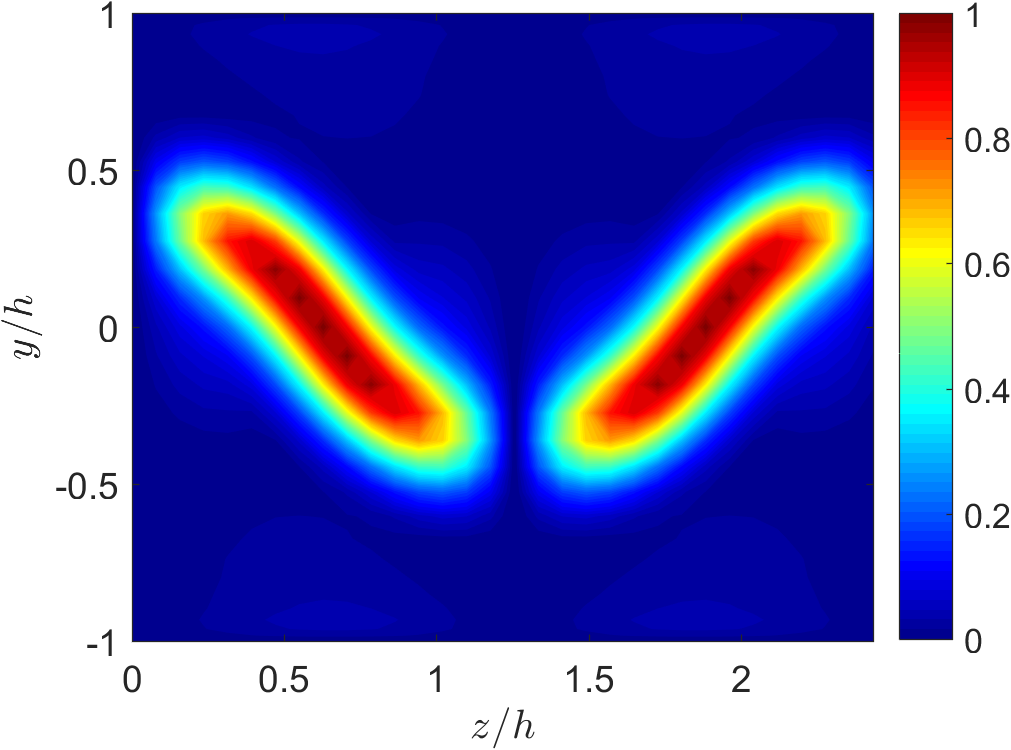}}
	\subfloat[]{\includegraphics[scale=0.32,trim=0.0cm 0cm 0.0cm 0cm, clip=true]{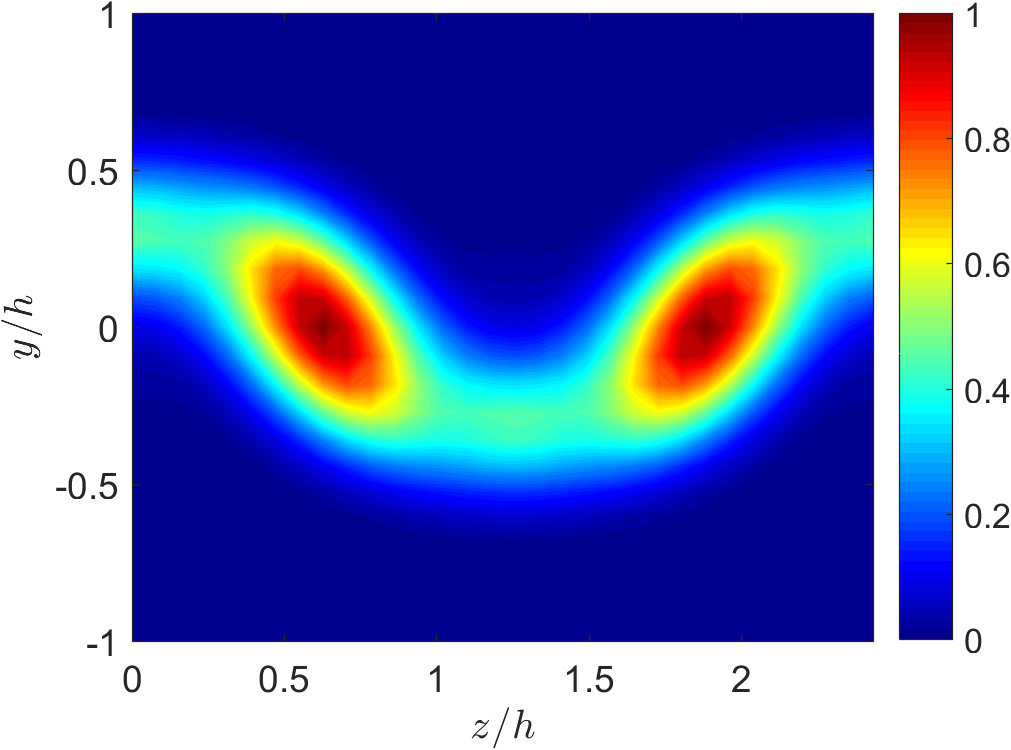}}
	
	\subfloat[]{\includegraphics[scale=0.32,trim=0cm 0cm 0.0cm 0cm, clip=true]{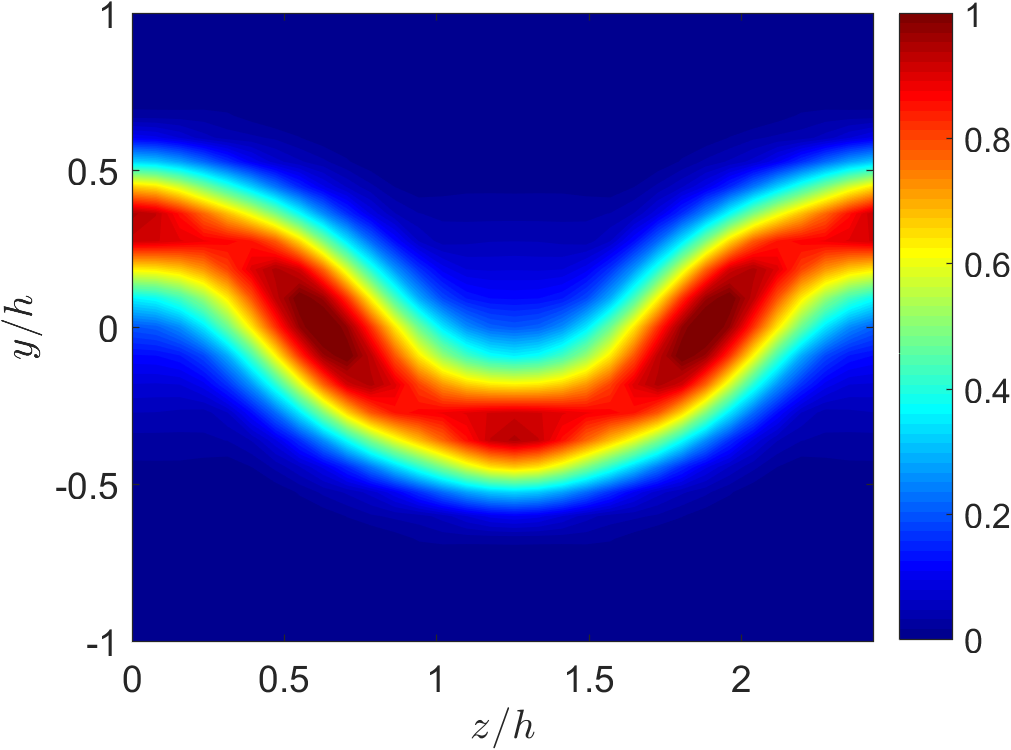}}
	\subfloat[]{\includegraphics[scale=0.32,trim=0.0cm 0cm 0.0cm 0cm, clip=true]{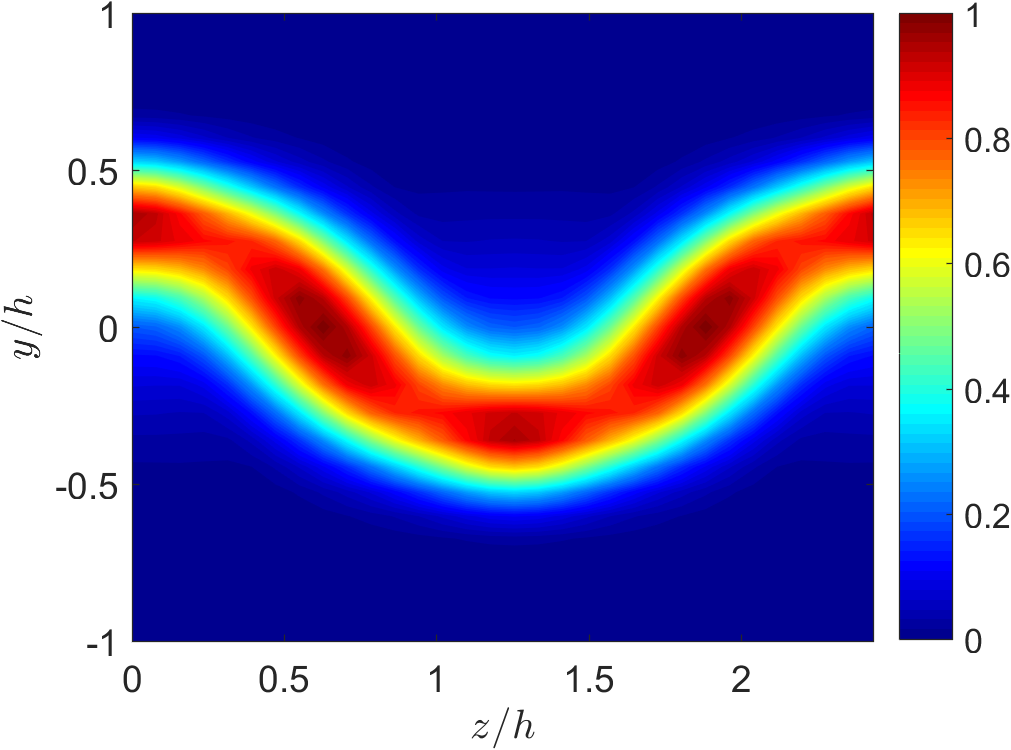}}
	\subfloat[]{\includegraphics[scale=0.32,trim=0.0cm 0cm 0.0cm 0cm, clip=true]{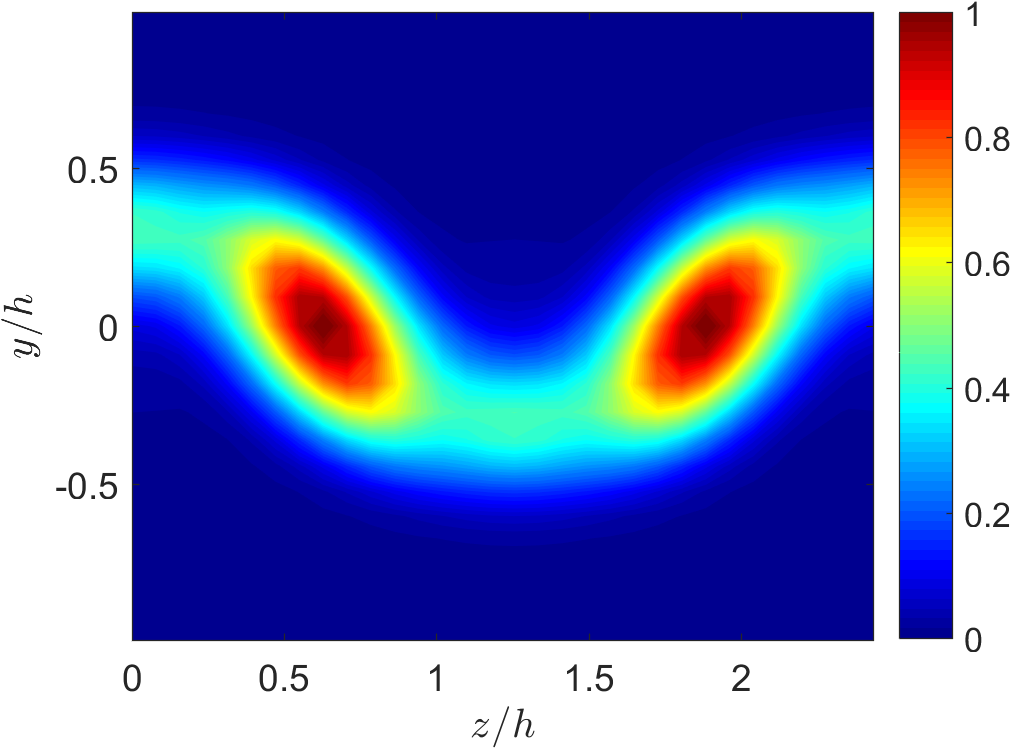}}
	\caption{\sean{Rows (1-3): cylinder results}. (a) The leading singular values computed \sean{for $\boldsymbol{k}_h$}, (b) the $u$-component of the optimal response mode $\hat{\boldsymbol{\psi}}_1(\sean{\boldsymbol{k}_h})$, and (c) the corresponding DMD mode. (d) The leading singular values computed \sean{for $2\boldsymbol{k}_h$}, (e) the $u$-component of the optimal response mode $\hat{\boldsymbol{\psi}}_1(2\sean{\boldsymbol{k}_h})$, and (f) the corresponding DMD mode. (g) The self-interaction of $\hat{\boldsymbol{\psi}}_1(\sean{\boldsymbol{k}_h})$  compared to the true nonlinear forcing $\hat{\boldsymbol{f}}(2\sean{\boldsymbol{k}_h})$ in (h), and (i) the forced resolvent mode ($u$- component) for $\sean{2\boldsymbol{k}_h}$. Rows (4-6): EQ1 results. As in (a)-(i) but for EQ1.}
	\label{fig:1st comparison}
\end{figure}

Let us denote $\boldsymbol{k}_h$ as the frequency/wavenumber of the first energetic harmonic for these two flows, i.e. the shedding frequeny $\omega_s$ and the fundamental streamwise wavenumber $k_{x_f}$ respectively. \citet{Symon18} showed that low-rank behavior for cylinder flow at $Re = 100$ is limited to a bandwidth of frequencies in the vicinity of $\boldsymbol{k}_h$. The singular values for $\boldsymbol{k}_h$ are plotted in Figure \ref{fig:1st comparison}(a), which shows $\sigma_1(\boldsymbol{k}_h)$ is two orders of magnitude larger than the rest. While progress can be made with this assumption, the forcing is, in fact, structured and influences the streamwise decay of the resolvent modes. This can be seen in Figure \ref{fig:1st comparison}(b-c), which compares the streamwise component of the resolvent and DMD modes. The overall agreement is good although the mode shapes begin to diverge around $x = 5$.

Similar observations can be made for EQ1. Figure \ref{fig:1st comparison}(j) indicates very low-rank behavior for $\boldsymbol{k}_h$ and, just like the cylinder case, the resolvent accurately predicts the mode shape as seen in Figure \ref{fig:1st comparison}(k-l) where the $u$-component of the first resolvent mode is compared to the true solution.

When the resolvent is not low-rank, such as at $\boldsymbol{k} =2\boldsymbol{k}_h$, as seen in Figure \ref{fig:1st comparison}(d,m), respectively, the leading mode from a \sean{SVD} is unlikely to predict the structure of the velocity fluctuations. Consequently, there is no agreement between the resolvent and DMD modes, which are plotted in Figure \ref{fig:1st comparison}(e-f), respectively, for the cylinder. The resolvent mode and true velocity fluctuations for EQ1 in Figure \ref{fig:1st comparison}(n-o), respectively, also disagree.

In both cases, there is no linear mechanism active at $2\boldsymbol{k}_h$. The analysis of \cite{Turton15} can be extended to argue which scale interactions are needed in Equation \ref{eq:approximate NLF} to recover $\hat{\boldsymbol{f}}(2\boldsymbol{k}_h)$. The linearized NSE are rewritten as
\begin{equation}
i q \sean{\boldsymbol{k}_h} \hat{\boldsymbol{u}}(q\boldsymbol{k}_h) = \boldsymbol{L}\hat{\boldsymbol{u}}(q\boldsymbol{k}_h) + \mathcal{N}(q\boldsymbol{k}_h),\label{eq:linearized NSE}
\end{equation}
\begin{equation} \mathcal{N}(q\boldsymbol{k}_h) = \sum_{r\neq 0, q} \mathcal{N}(\hat{\boldsymbol{u}}(r\boldsymbol{k}_h),\hat{\boldsymbol{u}}((q-r)\boldsymbol{k}_h).
\end{equation}
For $\| \hat{\boldsymbol{u}}(q\boldsymbol{k}_h) \| \sim \epsilon^{|q|}$, $\epsilon \ll 1$, as it must be for large singular values (Figures \ref{fig:1st comparison} (a,j)), and considering $q = 2$,
\begin{equation}
\underbrace{2i\sean{\boldsymbol{k}_h}\hat{\boldsymbol{u}}(2 \boldsymbol{k}_h)}_{\epsilon^2} = \underbrace{\boldsymbol{L}\hat{\boldsymbol{u}}(2\boldsymbol{k}_h)}_{\epsilon^2} + \underbrace{\mathcal{N}(\hat{\boldsymbol{u}}(\boldsymbol{k}_h),\hat{\boldsymbol{u}}(\boldsymbol{k}_h))}_{\epsilon^2} + \underbrace{\mathcal{N}(\hat{\boldsymbol{u}}(3\boldsymbol{k}_h),\hat{\boldsymbol{u}}(-\boldsymbol{k}_h))}_{\epsilon^4} +  \underbrace{\mathcal{N}(\hat{\boldsymbol{u}}(-\boldsymbol{k}_h),\hat{\boldsymbol{u}}(3\boldsymbol{k}_h))}_{\epsilon^4} + \cdots .
\end{equation}
Retaining terms on the order of $\epsilon^2$ results in
\begin{equation}
\hat{\boldsymbol{u}}(2\boldsymbol{k}_h) \approx \mathcal{H}(2\boldsymbol{k}_h)\mathcal{N}(\hat{\boldsymbol{u}}(\boldsymbol{k}_h),\hat{\boldsymbol{u}}(\boldsymbol{k}_h))\approx\hat{\boldsymbol{u}}(2\boldsymbol{k}_h) \approx -\sigma_1^2\chi_1^2\mathcal{H}(2\boldsymbol{k}_h)(\hat{\boldsymbol{\psi}}_1(\boldsymbol{k}_h) \cdot \nabla \hat{\boldsymbol{\psi}}_1(\boldsymbol{k}_h)),
\label{eq:2omega}
\end{equation}
where \sean{E}quation \ref{eq:rank1} is used to approximate $\hat{\boldsymbol{u}}(\boldsymbol{k}_h)$. \sean{Terms with the subscript 1 are associated with the wavenumber vector $\boldsymbol{k}_h$.}

Approximating the forcing $\hat{\boldsymbol{f}}(2\boldsymbol{k}_h)$ by the self-interaction of $\hat{\boldsymbol{\psi}}_1(\boldsymbol{k}_h)$ leads to a reasonable comparison with the true forcing for both the cylinder (Figures \ref{fig:1st comparison}(g-h)) and EQ1 (Figures \ref{fig:1st comparison}(p-q)). The modes excited by these forcings,  $\hat{\boldsymbol{u}}_f(2\boldsymbol{k}_h) = \mathcal{H}(2\boldsymbol{k}_h)\hat{\boldsymbol{f}}(2\boldsymbol{k}_h)$ show much improved agreement with the true response at $2\boldsymbol{k}_h$ (Figures \ref{fig:1st comparison}(i,r).)

The EQ1 structures in Figures \ref{fig:1st comparison}(r) and \ref{fig:1st comparison}(o) are nearly identical. While there is significant improvement in the mode shape for the cylinder, the agreement between Figure \ref{fig:1st comparison}(i) and the DMD mode in Figure \ref{fig:1st comparison}(f) is not perfect. The residual discrepancy for the cylinder is due to the fact $\hat{\boldsymbol{\psi}}_1(\sean{\boldsymbol{k}_h}) \neq \hat{\boldsymbol{u}}(\sean{\boldsymbol{k}_h})$ and only one triad was considered. Nevertheless, the forced resolvent mode captures the correct symmetries and is much more representative of the fluctuations at $2\sean{\boldsymbol{k}_h}$.

The mode at $2\boldsymbol{k}_h$ can be considered `parasitic' in that it is driven by the dominant linear mechanism at $\boldsymbol{k}_h$ and emerges from the resolvent operator being forced by the $\boldsymbol{k}_h$ velocity modes. It plays a role in the nonlinear forcing of the $\boldsymbol{k}_h$ mode. Setting $q = 1$ in Equation \ref{eq:linearized NSE} results in
\begin{equation}\label{eq:forcing_omega_s}
\underbrace{i\sean{\boldsymbol{k}_h}\hat{\boldsymbol{u}}(\boldsymbol{k}_h)}_{\epsilon} = \underbrace{\boldsymbol{L}\hat{\boldsymbol{u}}(\boldsymbol{k}_h)}_{\epsilon} + \underbrace{\mathcal{N}(\hat{\boldsymbol{u}}(2\boldsymbol{k}_h),\hat{\boldsymbol{u}}(-\boldsymbol{k}_h))}_{\epsilon^3} + \underbrace{\mathcal{N}(\hat{\boldsymbol{u}}(-\boldsymbol{k}_h),\hat{\boldsymbol{u}}(2\boldsymbol{k}_h))}_{\epsilon^3} +  \cdots .
\end{equation}
It was argued by \citet{Turton15} that the $\epsilon^3$ terms could be neglected resulting in $\boldsymbol{L}$ containing a marginally stable eigenvalue as discovered by \citet{Barkley06}. Retaining those terms and substituting equations \ref{eq:rank1} and \ref{eq:2omega}, we can express the forcing as
\begin{equation}\label{eq3_a}
\begin{split}
\hat{\boldsymbol{f}}(\boldsymbol{k}_h)
&  = \chi_1|\chi_1|^2\left \lbrace \left[\sigma_1^2\mathcal{H}(2\boldsymbol{k}_h)(\hat{\boldsymbol{\psi}}_1 \cdot \nabla \hat{\boldsymbol{\psi}}_1)\right] \cdot \nabla \left[\sigma_1\hat{\boldsymbol{\psi}}_1^*\right] + \left[\sigma_1\hat{\boldsymbol{\psi}}_1^*\right] \cdot \nabla \left[\sigma_1^2\mathcal{H}(2\sean{\boldsymbol{k}_h})(\hat{\boldsymbol{\psi}}_1 \cdot \nabla \hat{\boldsymbol{\psi}}_1)\right]\right \rbrace \\
& = \chi_1|\chi_1|^2\hat{\sean{\boldsymbol{f}}}'(\boldsymbol{k}_h),
\end{split}
\end{equation}
where notably $\hat{\sean{\boldsymbol{f}}}'(\boldsymbol{k}_h)$ can be computed from the resolvent with only knowledge of the mean.
From Equation \ref{eq:rank1} we can solve for the amplitude as
\begin{equation}
|\chi_1| = \left[\frac{1}{\left<\hat{\boldsymbol{\phi}}_1(\boldsymbol{k}_h),\hat{\boldsymbol{f}}'(\boldsymbol{k}_h)\right>}\right]^{1/2}.
\end{equation}

From \sean{E}quation \ref{eq:forcing_omega_s}, retaining terms at $\mathcal{O}(\epsilon)$ and treating the higher-order terms as an unknown forcing, the resolvent operator can be formulated and used to compute the spatial structure of the mode $\hat{\boldsymbol{\psi}}_1\sean{(\boldsymbol{k}_h)}$. Consequently, after cascading down the nonlinear interactions, at $\mathcal{O}(\epsilon^3)$ an equation can be derived for the corresponding amplitude of the mode, in close analogy to a weakly-nonlinear analysis. For EQ1 at $Re = 1000$, this approximation of $|\chi_1|$ has a relative error with respect to the true value computed directly from the solution of less than $1\%$. As \sean{E}quation \ref{eq3_a} and the subsequent calculation of $|\chi_1|$ relies on $\hat{\boldsymbol{\psi}}_1\sean{(\boldsymbol{k}_h)}$ closely approximating the true spatial structure, the agreement in the case of the cylinder is not as good. However if the DMD modes, which are the true velocity fluctuations at $\sean{\boldsymbol{k}_h}$ and $2 \sean{\boldsymbol{k}_h}$, are used, they can almost exactly reproduce $\hat{\boldsymbol{f}}(\sean{\boldsymbol{k}_h})$ \citep{Symon18b}, confirming the dominance of this interaction in the full flow. Moreover, if $\mathcal{H}(\sean{\boldsymbol{k}_h})$ is driven by this forcing, it results in a forced resolvent mode which very accurately approximates the DMD mode in Figure \ref{fig:1st comparison}(c). Thus it is possible to improve predictions from resolvent analysis, even when the operator is low-rank, by approximating the nonlinear forcing.

\begin{figure}
	\centerline{\includegraphics[scale=0.50]{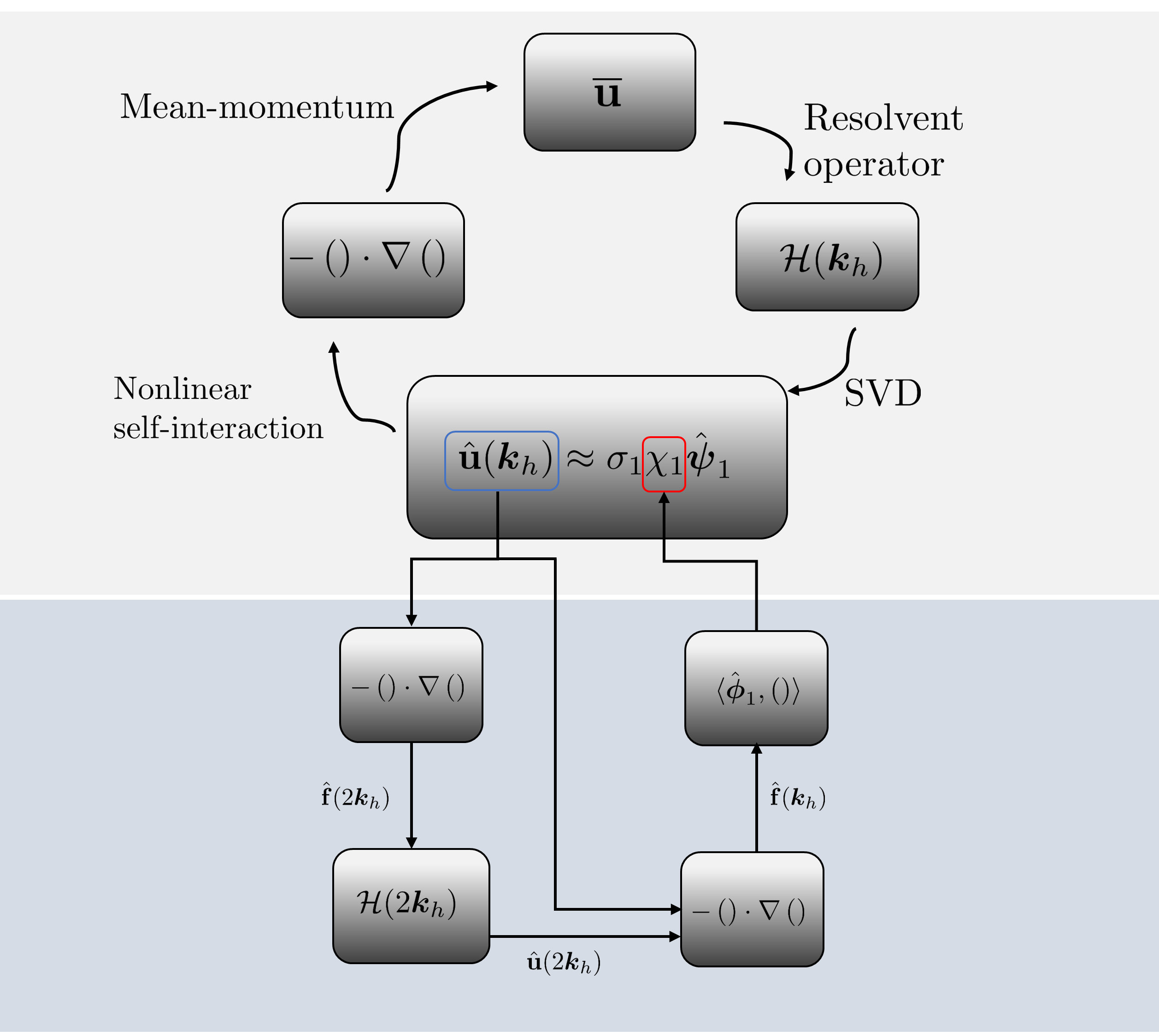}}
	\caption{The top of the schematic illustrates a variation of the self-sustaining process as viewed through the lens of resolvent analysis. The mean profile, fed into the resolvent operator, generates a highly amplified mode (obtained via the SVD) which self-interacts to generate the Reynolds stresses needed to sustain the mean and close the cycle. The saturated amplitude of the mode is obtained by considering the nonlinear interactions which give rise to higher harmonics and their subsequent feedback on the system, as shown in the bottom of the schematic.}
	\label{fig:ss_cycle}
\end{figure}

The results for the circular cylinder and EQ1 can be summarized in Figure \ref{fig:ss_cycle}. A SVD of the resolvent operator at $\boldsymbol{k}_h$ yields a highly amplified mode which interacts nonlinearly with its complex conjugate to produce the Reynolds stresses that sustain the mean profile. It also interacts with itself to produce the forcing at $2\boldsymbol{k}_h$. When $\mathcal{H}(2\boldsymbol{k}_h)$ is driven by this forcing, it results in a `parasitic' mode which nonlinearly interacts with the $-\boldsymbol{k}_h$ mode to produce the structured nonlinear forcing for $\boldsymbol{k}_h$. Thus in the case of the cylinder, where there is a greater mismatch between the resolvent mode and the DMD mode, the original $\boldsymbol{\psi}_1(\sean{\boldsymbol{k}_h})$ can be iteratively improved using the feedback loop in Figure \ref{fig:ss_cycle}.

\section{Summary}
Resolvent analysis was used to identify the linear mechanisms and dominant nonlinear interactions required to provide a self-consistent description of low-Reynolds number cylinder flow and the Couette equilibrium solution EQ1. It was demonstrated that the leading resolvent mode, obtained by approximation of the resolvent in a region where it is low-rank, is able to correctly capture the spatial structure of the most energetic mode. Considering the relevant scale interactions, i.e. approximating the forcing and leveraging the low rank nature of the energetic mode generates improved approximations of the second harmonic, here termed a parasitic mode because of its link to the interactions of the energetic mode and the lack of low-rank behavior of its resolvent. These ideas can be consolidated to provide a means to determine the amplitude of the dominant energetic mode, and hence effectively close the system, using knowledge of only the mean profile, in cases where the resolvent is not low-rank for all scales. It is hypothesized that this framework may also have relevance for flows with multiple linearly dominant energetic modes. The hope would be to use the resolvent operator to obtain low-dimensional representations of the important linearly-amplified structures, which are hypothesized to correspond to identifiable physical mechanisms, and use the notion of parasitic modes to provide an accurate representation of the weakly energetic but nonlinearly relevant scales.

\vspace{0.2in}
The support of ONR under grants N00014-17-1-2307 and N00014-17-1-3022 is gratefully acknowledged.

\bibliographystyle{authordate1}
\bibliography{DMDparasitic}

\end{document}